\documentclass[runningheads]{llncs}
\usepackage[T1]{fontenc}
\usepackage{graphicx}
\usepackage{amsmath}
\usepackage{amsfonts}
\begin{document}
\title{Multilingual TASU: Extending Text-Only Alignment to Multilingual Speech Recognition via Simulated CTC Posteriors}
\titlerunning{Multilingual TASU}
% \author{Anonymous Author(s)}
% \authorrunning{Anonymous Author(s)}
% \institute{Anonymous Institution}
%
\author{%
Chenghao Wang\inst{2,*},
Yi Yang\inst{1,*},
Jing Peng\inst{1},
Yu Xi\inst{1},
Shuai Wang\inst{3},
Kai~Yu\inst{1,**}
}

\authorrunning{C. Wang et al.}

\institute{%
\resizebox{\textwidth}{!}{%
\begin{tabular}{@{}c@{}}
\textsuperscript{1}X-LANCE Lab, Department of Computer Science and Engineering, Shanghai Jiao Tong University\\
\textsuperscript{1}MoE Key Lab of Artificial Intelligence,
\quad
\textsuperscript{1}Jiangsu Key Lab of Language Computing, China\\
\textsuperscript{2}AISpeech Co., Ltd., Suzhou, China\\
\textsuperscript{3}Nanjing University, China\\[0.8ex]
\texttt{\{jing.peng,kai.yu\}@sjtu.edu.cn, voldbaboon@gmail.com, shuaiwang@nju.edu.cn}
\end{tabular}%
}%
}

\maketitle              % typeset the header of the contribution
\begingroup
\renewcommand{\thefootnote}{*}
\footnotetext[1]{These authors contributed equally.}

\renewcommand{\thefootnote}{**}
\footnotetext[2]{Indicates the corresponding author.}
\endgroup

\begin{abstract}
Speech Large Language Models (Speech LLMs) achieve strong multilingual ASR when large-scale paired speech-text data is available, yet such resources remain scarce and highly imbalanced across languages. Text-Only Alignment for Speech Understanding (TASU) alleviates this bottleneck by training cross-modal alignment from simulated CTC posteriors without paired audio, but prior work has been limited to English, leaving open whether text-driven alignment can scale to the multilingual setting where data sparsity and cross-lingual phonetic interference amplify the difficulty of learning a shared alignment interface.
In this work, we investigate TASU-style posterior alignment as a means of mitigating paired-data scarcity in multilingual ASR. We adopt the native 50,002-way posterior space of a frozen OWSM-CTC frontend as a shared multilingual interface to Qwen2.5-1.5B-Instruct, and systematically compare three supervision sources on eight Multilingual LibriSpeech languages: real audio-derived posteriors, heuristic CTC Posterior Simulation (CPS), and posteriors generated by a learned multilingual simulator. 
With real posteriors, LoRA adaptation reduces pooled WER from 10.08\% (standalone OWSM-CTC) to 8.21\%, improving all eight languages. For text-only alignment, the learned simulator substantially outperforms heuristic CPS and attains 10.51\% WER with LoRA. Notably, a CPS-initialized model reaches 9.19\% WER after only one real-posterior adaptation epoch using merely one-ninth of the complete MLS paired-speech training set, demonstrating that text-first initialization can drastically reduce the amount of paired audio required for competitive multilingual ASR. These findings establish TASU as a viable multilingual alignment framework and highlight text-only simulation as an effective remedy for paired-data scarcity and imbalance across languages.

\end{abstract}
\section{Introduction}
Speech Large Language Models (Speech LLMs) extend large language models to speech processing by aligning speech representations with the LLM's input space, achieving state-of-the-art results in speech recognition and understanding domains~\cite{arora2025landscape,peng2024survey}. 
The dominant alignment paradigm relies on continuous feature projection trained with paired audio--text data~\cite{tang2023salmonn,chu2024qwen2,ma2024embarrassingly,xu2025qwen2,ding2025kimi}, which faces severe scalability constraints in the multilingual setting: paired speech--text resources are not only scarce but also heavily imbalanced across languages, leaving many languages with insufficient data to train a dedicated alignment module.

To reduce this dependence, recent work explores connectionist temporal classification (CTC)-based alignment strategies~\cite{graves2006connectionist}. 
Text-Only Alignment for Speech Understanding (TASU)~\cite{peng2026tasu} and TASU2~\cite{peng2026tasu2} simulate CTC posterior distributions from text transcripts via CTC Posterior Simulation (CPS), enabling projector training without paired audio. 
After compacting via Label-Synchronous Decoding (LSD)~\cite{chen2016phone,chen2019lsd}, these posteriors can be approximated from one-hot vectors, achieving competitive recognition and multitask generalization in English~\cite{peng2026tasu}. 
TASU2 further introduces a simulator controllable by word error rate (WER) for improved alignment fidelity and cross-domain transfer~\cite{peng2026tasu2}.

However, both TASU and TASU2 have been evaluated primarily in English. Extending posterior-based alignment to multilingual speech recognition raises three questions: (i) Can the native posterior space of a multilingual CTC encoder serve as a
shared alignment interface across languages? (ii) How closely can heuristic and learned text-conditioned posteriors
approach supervision from real audio-derived posteriors? (iii) Can text initialization be combined with limited paired speech for effective multilingual adaptation?

This paper investigates these questions using OWSM-CTC~\cite{peng2024owsmctc}---an encoder-only model trained on 180{,}000 hours of audio across 151 languages---as the frozen multilingual posterior provider. We retain its native 50,002-way CTC posterior space and use the same projector and language model for all languages. We evaluate eight languages from Multilingual LibriSpeech (MLS)~\cite{pratap2020mls}---English, German, Dutch, French, Spanish, Italian, Portuguese, and Polish---under four settings: standalone OWSM-CTC, TASU trained with heuristic CPS, TASU trained with posteriors generated by a learned multilingual simulator, and TASU trained with real audio-derived OWSM posteriors. We further examine projector-only and low-rank adaptation (LoRA) settings, as well as a reduced-paired-speech setting that initializes from CPS training before real-posterior adaptation.
Besides, to verify that these findings are not specific to the OWSM-CTC encoder, we additionally replicate the key configurations with a Whisper-CTC frontend whose CTC head is trained on limited data, confirming that the TASU curriculum-learning gains generalize across different posterior providers.

Our results demonstrate that the native multilingual CTC posterior space provides an effective shared interface for Speech LLM alignment. LoRA adaptation with real posteriors consistently improves recognition across all eight languages and outperforms the standalone OWSM-CTC baseline. Learned simulation substantially narrows the gap between heuristic CPS and audio-derived supervision, and approaches the OWSM-CTC baseline when combined with LoRA, although it remains behind real-posterior training. In addition, initializing TASU with text-derived posteriors before adaptation with a limited amount of paired speech provides a practical reduced-resource training strategy. Overall, the results show that multilingual TASU is feasible with a shared posterior interface, while the gap between simulated and real posterior supervision remains an important limitation of text-conditioned supervision.

In summary, our contributions are:
\begin{itemize}
    \item We extend TASU to a single shared eight-language Speech LLM using the
    native multilingual posterior space of OWSM-CTC, without introducing
    language-specific alignment modules.
    \item We conduct a controlled comparison of heuristic CPS, learned
    simulated posteriors, and real audio-derived posteriors under both
    projector-only and parameter-efficient adaptation, quantifying the effect
    of posterior source on multilingual alignment.
    \item We examine multilingual TASU beyond full-data training through a
    reduced-paired-speech adaptation setting and a simulator language-coverage
    ablation, clarifying both the opportunities and limitations of
    text-derived posterior supervision.
    \item We validate the encoder-generalization capability of multilingual
    TASU by replicating the core training configurations with an alternative
    Whisper-CTC frontend, demonstrating consistent curriculum-learning gains
    across different CTC posterior providers.
\end{itemize}

\section{Related Work}
\label{sec:related_work}

\subsection{Speech-LLM Alignment and CTC Compact Representations}
\label{subsec:rw_alignment_ctc}

Speech LLMs align speech and text through either continuous feature projection---which preserves acoustic detail but introduces redundancy and demands large paired datasets~\cite{tang2023salmonn,chu2024qwen2,ma2024embarrassingly}---or CTC-based compact representations~\cite{ma2025legoslm}. CTC~\cite{graves2006connectionist} defines frame-level posteriors over output symbols and a blank token; its collapse operation removes blanks and merges repeated labels. LSD~\cite{chen2016phone,chen2017ctc,chen2019lsd} further compresses these posteriors by discarding blank-dominated frames and merging consecutive identical tokens, yielding representations structurally closer to text embeddings. This property motivates using CTC posteriors as the alignment interface for speech-text modalities.

\subsection{Text-Only Alignment: TASU and TASU2}
\label{subsec:rw_tasu}

\paragraph{TASU.}
Text-Only Alignment for Speech Understanding (TASU)~\cite{peng2026tasu}
aligns speech and text at the CTC posterior level rather than directly
projecting acoustic encoder hidden states. Its central idea is to construct a
shared $V$-dimensional posterior interface for alignment whose inputs can be obtained from
text during training and from speech during inference.

During text-only training, a transcript
$\mathbf{y}=(y_1,\ldots,y_U)$ is tokenized using the vocabulary associated
with the CTC encoder. CTC Posterior Simulation (CPS) first converts each token
$y_u$ into a smoothed one-hot posterior:
\begin{equation}
\tilde{\mathbf{p}}_u
=
\alpha\boldsymbol{\delta}_{y_u}
+
(1-\alpha)\frac{1}{V}\mathbf{1},
\qquad
\alpha\sim\mathcal{U}
(\lambda_{\mathrm{low}},\lambda_{\mathrm{high}}),
\label{eq:cps_smooth}
\end{equation}
where one confidence value $\alpha$ is sampled for the sequence.
CPS then independently deletes posterior frames with probability
$p_{\mathrm{del}}$ and inserts additional frames according to
$p_{\mathrm{ins}}$. Each inserted frame is either a duplicate of a neighboring
posterior or a one-hot blank vector. These operations imitate token omissions,
repetitions, and blank insertions commonly observed in CTC output sequences.
The resulting pseudo-posterior sequence is mapped by a trainable projector
into the embedding space of a frozen LLM, which predicts the transcript under
teacher forcing. Thus, neither waveform input nor execution of the acoustic
encoder is required during this alignment stage, and the original TASU setup
optimizes only the projector.

At inference time, the simulated posterior sequence is replaced by real
posteriors $\mathbf{P}\in\mathbb{R}^{T\times V}$ obtained by applying the
frozen CTC encoder to speech. Because frame-level CTC outputs contain many
blank-dominated frames and repeated labels, TASU applies Label-Synchronous
Decoding (LSD)~\cite{chen2016phone,chen2019lsd}. Frames whose blank
probability exceeds a threshold $\tau$ are removed:
\begin{equation}
\mathbf{P}_t'
=
\begin{cases}
\emptyset, & P_{t,\mathrm{blank}}>\tau,\\
\mathbf{P}_t, & \text{otherwise},
\end{cases}
\label{eq:lsd_removal}
\end{equation}
after which consecutive retained frames sharing the same non-blank argmax
label are averaged:
\begin{equation}
\mathbf{P}_j''
=
\frac{1}{|S_j|}
\sum_{t\in S_j}\mathbf{P}_t',
\qquad j=1,\ldots,J,
\label{eq:lsd_merge}
\end{equation}
where $S_j$ is the $j$-th group of consecutive frames and $J$ is the
compressed sequence length. The same projector trained with CPS then maps
these real compressed posteriors into the LLM. TASU therefore transfers from
text-only training to speech inference by making the simulated and
audio-derived inputs share the same CTC posterior space. TASU can also serve
as the first stage of a curriculum: CPS establishes an initial
posterior-to-LLM alignment from transcripts, followed by adaptation with a
smaller amount of real audio-derived supervision.

\paragraph{TASU2.}
The heuristic CPS procedure in TASU constructs pseudo-posteriors through
random smoothing, deletion, and insertion, but does not learn the mapping
from text to CTC posterior sequences. TASU2~\cite{peng2026tasu2} replaces
this procedure with an autoregressive Transformer simulator that predicts a
sequence of posterior distributions from tokenized text. In its original
formulation, the simulator is additionally conditioned on a discrete WER code,
allowing posterior generation to be controlled across different error
regimes.

Given a transcript $\mathbf{y}$, the simulator predicts
$\hat{\mathbf{P}}=(\hat{\mathbf{p}}_1,\ldots,\hat{\mathbf{p}}_{\hat{T}})$
and is trained to match the corresponding teacher CTC posterior sequence
$\mathbf{P}=(\mathbf{p}_1,\ldots,\mathbf{p}_{T})$. Variable-length targets
are padded to $T_{\mathrm{train}}$, and posterior-level cross-entropy is
computed only over valid frames:
\begin{equation}
\mathcal{L}_{\mathrm{sim}}(\theta)
=
\mathbb{E}
\left[
-\frac{1}{\sum_t m_t}
\sum_{t=1}^{T_{\mathrm{train}}}m_t
\sum_{v=1}^{V}
p_{t,v}\log\hat{p}_{t,v}
\right],
\label{eq:sim_ce}
\end{equation}
where $m_t$ equals one at valid target-posterior positions and zero at padded
positions. Compared with heuristic CPS, this learned formulation can model
posterior duration, token confusions, blank patterns, and contextual
dependencies directly from teacher posteriors.

Our multilingual setting adopts this learned text-to-posterior formulation
but does not use the WER-control data construction of the original TASU2.
Specifically, we do not perform waveform augmentation or divide training
examples into WER intervals. Instead, the simulator directly learns to map
multilingual OWSM-tokenized transcripts to their corresponding OWSM-CTC
posterior sequences under a single fixed condition. Once trained, it is
frozen and used to generate posterior supervision from multilingual text for
downstream TASU training.

\subsection{Multilingual Speech Recognition}
\label{subsec:rw_multilingual}

Multilingual ASR requires handling phonetic diversity under shared representations. OWSM-CTC~\cite{peng2024owsmctc}---an encoder-only model trained on 180{,}000 hours across 151 languages---demonstrates competitive multilingual recognition with efficient inference. We leverage OWSM-CTC as the frozen posterior provider and evaluate on MLS~\cite{pratap2020mls} to investigate how well simulated CTC posteriors can approximate audio-derived supervision in multilingual settings.

\section{Experimental Setup}
\label{sec:experimental_setup}

\begin{figure}[t]
\centering
\includegraphics[width=\linewidth]{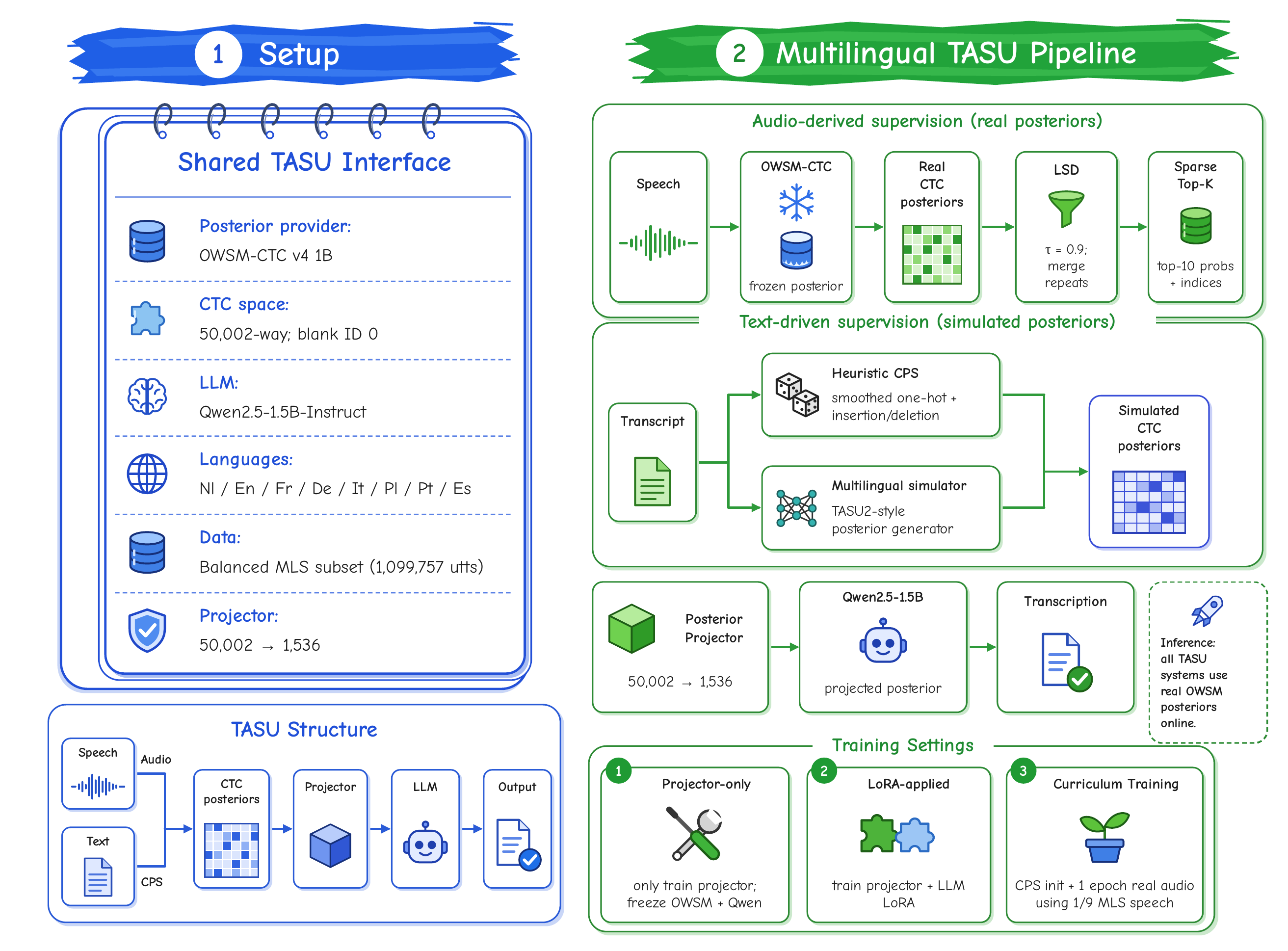}
\caption{Overview of the multilingual TASU framework. The left panel
summarizes the shared model interface. The right panel shows the
audio-derived and text-driven posterior-generation paths, followed by the
common projector and language model. We evaluate projector-only training,
LoRA adaptation, and CPS initialization followed by reduced-speech
real-posterior adaptation. All systems use real OWSM-CTC posteriors at
inference.}
\label{fig:multilingual_tasu_overview}
\end{figure}

Figure~\ref{fig:multilingual_tasu_overview} provides an overview of the
multilingual TASU setup and the evaluated training strategies. All systems
share the same OWSM-CTC posterior space, posterior projector, and language
model. They differ in the source of posterior supervision used during
training and in whether the language model is adapted with LoRA. At inference,
all systems receive real OWSM-CTC posteriors extracted from input speech.

\subsection{Multilingual TASU Formulation}

For an input posterior sequence $\mathbf{P}$ and transcript
$\mathbf{y}=(y_1,\ldots,y_N)$, the projector maps each posterior frame to
the Qwen embedding space, producing $\mathbf{Z}=(\mathbf{z}_1,\ldots,
\mathbf{z}_T)$. The projected sequence replaces the \texttt{<speech>}
placeholder in the instruction prompt. Under teacher forcing, TASU is
optimized with

\begin{equation}
\mathcal{L}_{\mathrm{TASU}}
=
-\frac{1}{N+1}
\sum_{n=1}^{N+1}
\log p_{\phi}(y_n \mid \mathbf{q},\mathbf{Z},y_{<n}),
\end{equation}

where $\mathbf{q}$ denotes the textual instruction and $y_{N+1}$ is the
end-of-sequence token. Prompt and posterior positions are excluded from
the loss. All eight languages share the same posterior projector, Qwen
backbone, and LoRA parameters; no language-specific projector or output
head is introduced.

\subsection{Dataset}
\label{subsec:dataset}

We conduct all experiments on eight languages from Multilingual LibriSpeech (MLS)~\cite{pratap2020mls}: Dutch, English, French, German, Italian, Polish, Portuguese, and Spanish. The original MLS training data are highly imbalanced, with substantially more utterances available for Dutch, English, French, German, and Spanish than for Italian, Polish, and Portuguese. To obtain a more balanced training set, we cap each higher-resource language at 196,000 utterances while retaining nearly all available data from the three lower-resource languages. This sampling procedure produces approximately 1.1 million utterances, or about one third of the complete MLS training collection. The fraction is thus a consequence of language balancing rather than uniform subsampling. Table~\ref{tab:train_data} reports the resulting distribution.

\begin{table}[t]
\centering
\caption{Per-language training-utterance counts in the complete MLS corpus and the subset used in this work.}
\label{tab:train_data}
\small
\setlength{\tabcolsep}{5.5pt}
\begin{tabular}{@{}lrr@{\hspace{2em}}lrr@{}}
\hline
\textbf{Language} & \textbf{Full MLS} & \textbf{Used}
& \textbf{Language} & \textbf{Full MLS} & \textbf{Used} \\
\hline
Dutch     & 374,287   & 196,000 & Italian    & 59,623  & 58,431 \\
English   & 2,420,228 & 196,000 & Polish     & 25,043  & 24,543 \\
French    & 258,213   & 196,000 & Portuguese & 37,533  & 36,783 \\
German    & 469,942   & 196,000 & Spanish    & 220,701 & 196,000 \\
\hline
\multicolumn{2}{r}{\textbf{Full MLS total}} & 3,865,570
& \multicolumn{2}{r}{\textbf{Used total}} & 1,099,757 \\
\hline
\end{tabular}
\end{table}

All training conditions are derived from this balanced data pool. The real-posterior condition uses OWSM-CTC posteriors extracted from the corresponding speech, whereas heuristic CPS uses only the transcripts. For the learned-simulator condition, we apply a lightweight quality filter to remove a small number of clearly pathological posterior sequences, including extreme-length and blank-dominated outputs. In the reduced-paired-speech experiment, text initialization uses the complete balanced transcript set, followed by a single real-posterior adaptation epoch on one third of the corresponding speech subset, equivalent to approximately one ninth of the complete MLS training collection. Development and evaluation use the official
MLS splits for all eight languages.

\subsection{Model Configuration}
\label{subsec:model_configuration}

Table~\ref{tab:model_configuration} summarizes the common model configuration. We use the frozen OWSM-CTC v4 1B model~\cite{peng2024owsmctc} as the multilingual acoustic posterior provider. Its 50,002 CTC output classes include the blank class at index 0. Each utterance is processed with the corresponding OWSM language symbol and the ASR task symbol. We then apply label-synchronous posterior
compression with a blank threshold of 0.9, averaging consecutive frames that share the same non-blank argmax label.

Because OWSM-CTC remains frozen throughout TASU training, we precompute its audio-derived posteriors once rather than repeatedly running the acoustic frontend at every training iteration. CTC posteriors are typically highly concentrated, allowing us to retain only the ten largest probabilities and their token indices at each compressed frame. The projector operates directly on this sparse representation, substantially reducing both data-loading overhead and the computation associated with the 50,002-dimensional input. The retained values are the original OWSM probabilities and are not renormalized; probabilities outside the top ten are treated as zero. This offline representation is used only for training, while evaluation computes OWSM-CTC posteriors directly from the input speech.

\begin{table}[t]
\centering
\caption{Common model and posterior configuration.}
\label{tab:model_configuration}
\small
\renewcommand{\arraystretch}{1.12}
\begin{tabular}{@{}l@{\hspace{1em}}p{0.67\linewidth}@{}}
\hline
\textbf{Component} & \textbf{Configuration} \\
\hline
Posterior provider & Frozen OWSM-CTC v4 1B \\
CTC interface & Vocabulary size 50,002; blank ID 0 \\
Compression & Blank threshold $\tau=0.9$; consecutive-label averaging \\
Sparse storage & Top-$K=10$ probabilities and indices; no renormalization \\
Projector & Single affine map, $50{,}002 \rightarrow 1{,}536$; no bottleneck \\
Language model & Qwen2.5-1.5B-Instruct~\cite{qwen25}; hidden size 1,536 \\
Heuristic CPS & $\lambda \sim \mathcal{U}(0.8,1.0)$; $p_{\mathrm{del}}=p_{\mathrm{ins}}=0.05$ \\
Simulator architecture & Transformer; $d_{\mathrm{model}}=512$; 8 heads; 6 encoder and 6 decoder layers; feed-forward size 2,048 \\
Simulation setting & WER bucket 1; temperature 0.5 \\
LoRA~\cite{hu2022lora} & $r=64$; $\alpha=16$; dropout 0.05; attention and feed-forward projections \\
\hline
\end{tabular}
\end{table}

The compressed posterior is mapped directly into the input space of Qwen2.5-1.5B-Instruct by a trainable projector. For a sparse frame with retained indices $i_k$ and values $p_{t,i_k}$, the implementation
computes
\begin{equation}
    \mathbf{z}_t = \sum_{k=1}^{10} p_{t,i_k}\mathbf{W}_{i_k}+\mathbf{b},
\label{eq:sparse_projection}
\end{equation}
which is exactly equivalent to applying the dense affine layer after scattering the retained values into a 50,002-dimensional zero vector. The projected sequence replaces a single \texttt{<speech>} placeholder in the
Qwen prompt. 

For the encoder-generalization experiments (Section~\ref{subsec:whisper_ctc}), we use Whisper-Large-v3~\cite{radford2023whisper} with a randomly initialized CTC head aligned to the Qwen2.5-1.5B-Instruct vocabulary (151,667 classes). Because limited training data is insufficient for this large vocabulary, we additionally evaluate a reduced 38,243-class head. The CTC head is trained on the same balanced MLS subset used in the OWSM-CTC set.

The CTC head is trained with AdamW~\cite{loshchilov2019adamw} at a peak learning rate of $5\times10^{-4}$, using a 1{,}000-step warm-up and cosine decay. Training uses bfloat16 and DeepSpeed ZeRO-2~\cite{rajbhandari2020zero} on eight Ascend NPUs, with dynamic batching under a frame budget of 3{,}200 and no gradient accumulation. We trained for three epochs with random seed 42.

\subsection{Experimental Configurations}
\label{subsec:experimental_configurations}

We organize the experiments according to the source of the posterior used during TASU adaptation.  \emph{Real} uses offline posteriors extracted from the paired MLS audio by OWSM-CTC.  \emph{CPS} constructs heuristic pseudo-posteriors directly from transcripts.  \emph{Simulator} uses a frozen multilingual TASU2 simulator to predict posteriors from OWSM-tokenized text. The simulator is trained separately from real teacher posteriors; therefore, ``text-only'' in this condition refers to the downstream TASU adaptation stage rather than the complete simulator-training pipeline.  At evaluation time, every TASU system receives real posteriors produced online by the same OWSM-CTC frontend.

In the projector-only conditions, the OWSM frontend and Qwen backbone are frozen and only the posterior projector is optimized.  In the LoRA conditions, the projector remains trainable and LoRA parameters are added to the Qwen attention projections (\texttt{q}, \texttt{k}, \texttt{v}, and \texttt{o}) and feed-forward projections (\texttt{gate}, \texttt{up}, and \texttt{down}).  The two-stage CPS$\rightarrow$Real condition initializes from the text-trained checkpoint and performs one additional epoch using the reduced real-posterior subset.  Finally, the simulator-coverage ablation removes Polish teacher-posterior pairs only from simulator training. The resulting seven-language simulator generates posteriors for all eight languages, and the downstream TASU model remains trained on all eight languages.

In addition to the main experiments, we replicate a subset of these configurations: real-posterior (audio-only) training, CPS-only training, and the CPS $\rightarrow$ Real curriculum with an alternative Whisper-CTC frontend to verify that the TASU gains are not specific to OWSM-CTC. The learned simulator is not included in this replication because it is trained to generate OWSM-CTC posteriors and does not directly apply to the Whisper-CTC output space.

\subsection{Training and Evaluation}
\label{subsec:training_evaluation}

Models are trained with AdamW~\cite{loshchilov2019adamw} at a peak learning rate of $5\times10^{-5}$,
using a 700-step warm-up followed by cosine decay. Training uses bfloat16 and DeepSpeed ZeRO-2~\cite{rajbhandari2020zero} across 16 Ascend NPUs. Each optimizer step processes one dynamically constructed pack per device, containing at most four utterances subject to the configured length budget. No gradient accumulation is used. Models are trained for up to three epochs with a maximum of 55,000 optimizer steps. All runs use random seed 42. Offline real and simulated posteriors share the same sparse representation and data-loading path, ensuring that their downstream projector computation is identical.

For the supplementary encoder-generalization experiments, we instantiate an alternative
frontend by initializing the acoustic encoder from Whisper-Large-v3 and
training a randomly initialized CTC head on the same balanced MLS subset.
The CTC head vocabulary is aligned with Qwen2.5-1.5B-Instruct, yielding a
150k-way output layer. Because limited training data is insufficient
to reliably estimate parameters for this large vocabulary, we
additionally evaluate a reduced 38k-vocabulary CTC head. All
downstream TASU training and evaluation configurations---including
optimizer, learning rate, LoRA hyperparameters, and decoding protocol---are kept identical to the OWSM-CTC experiments to ensure a controlled comparison.

We decode each official MLS test split with its corresponding OWSM language symbol. References and hypotheses are Unicode-normalized, lowercased, and
stripped of ASR punctuation while retaining intra-word apostrophes. We report WER using the WeNet scoring implementation~\cite{zhang2022wenet}.
The \emph{Overall} column is obtained by pooling all reference words across the eight test sets rather than taking an unweighted mean of the eight language-level WERs.

\section{Results and Analysis}
\label{sec:results}

\subsection{Multilingual TASU with OWSM-CTC}
\label{subsec:owsm_ctc_results}
\paragraph{Posterior-source comparison with a frozen LLM.} Table~\ref{tab:projector_results} compares posterior sources when only the posterior projector is trained. Training with real OWSM posteriors obtains almost the same WER as the original OWSM-CTC, and the gain is not uniform: French and Portuguese improve substantially, whereas Polish degrades from 15.09 to 23.55. Training with learned simulated posteriors obtains 11.50 WER and is considerably more effective than the heuristic CPS configuration at 20.55, but neither simulated condition matches real-posterior training. These results show that the shared multilingual posterior interface is viable, while the mismatch between text-derived and real posteriors remains a central limitation, and training only the projector between the CTC posteriors and the LLM may not be sufficient.

\begin{table}[t]
\centering
\caption{WER (\%) with the Qwen backbone frozen and only the posterior projector trained. ``Sim.'' denotes the learned multilingual simulator. Nl, En, Fr, De, It, Pl, Pt, and Es denote Dutch, English, French, German, Italian, Polish, Portuguese, and Spanish, respectively; Overall denotes pooled WER.}
\label{tab:projector_results}
\resizebox{\textwidth}{!}{%
\begin{tabular}{llrrrrrrrrr}
\hline
System & Training posterior & Nl & En & Fr & De & It & Pl & Pt & Es & Overall \\
\hline
OWSM-CTC       & --   & 15.04 & \textbf{7.03} & 10.17 & \textbf{8.27} & \textbf{15.65} & \textbf{15.09} & 12.95 & \textbf{5.83} & 10.08 \\
TASU-CPS       & CPS  & 15.66 & 28.69 & 24.63 & 13.38 & 22.65 & 33.98 & 21.66 & 15.60 & 20.55 \\
TASU-Sim.      & Sim. & 18.58 & 7.82 & 8.70 & 9.48 & 18.19 & 23.92 & 12.08 & 7.44 & 11.50 \\
TASU-Real      & Real & \textbf{14.77} & 7.07 & \textbf{7.82} & 8.58 & 16.32 & 23.55 & \textbf{10.58} & 5.97 & \textbf{9.93} \\
\hline
\end{tabular}}
\end{table}

\paragraph{Parameter-efficient multilingual adaptation.}
Table~\ref{tab:lora_results} reports the conditions in which the shared Qwen backbone is adapted with LoRA in addition to training the posterior projector. With real posteriors, TASU-Real-LoRA reaches 8.21 overall WER, an 18.6\% relative reduction from OWSM-CTC. Unlike projector-only real-posterior training, it improves all eight languages, with particularly large absolute gains for French, Italian, and Portuguese.  LoRA also improves TASU trained with learned simulated posteriors from 11.50 to 10.51 overall WER, although it remains behind both OWSM-CTC and real-posterior adaptation.  Thus, adapting the language model helps exploit both posterior sources, but does not eliminate the mismatch between learned simulated and real acoustic posteriors.

\begin{table}[t]
\centering
\caption{WER (\%) for LoRA-based TASU adaptation. CPS$\rightarrow$Real uses only one ninth of the complete MLS paired-speech training set in a single real-posterior adaptation epoch.}
\label{tab:lora_results}
\resizebox{\textwidth}{!}{%
\begin{tabular}{llrrrrrrrrr}
\hline
System & Adaptation data & Nl & En & Fr & De & It & Pl & Pt & Es & Overall \\
\hline
OWSM-CTC                  & --              & 15.04 & 7.03 & 10.17 & 8.27 & 15.65 & 15.09 & 12.95 & 5.83 & 10.08 \\
TASU-Sim.-LoRA            & Simulated       & 18.53 & 7.82 & 7.62 & 7.77 & 15.45 & 16.38 & 12.04 & 6.31 & 10.51 \\
TASU-CPS$\rightarrow$Real-LoRA & Text $+$ 1/9 real & 14.12 & 6.73 & 7.18 & 7.94 & 14.85 & 18.32 & 9.47 & 5.54 & 9.19 \\
TASU-Real-LoRA (one epoch)           & 1/9 real          & 15.26 & 8.67 & 8.50 & 9.83 & 17.78 & 26.76 & 12.25 & 6.70 & 11.04 \\
TASU-Real-LoRA            & 1/3 real        & \textbf{12.96} & \textbf{6.50} & \textbf{6.48} & \textbf{6.68} & \textbf{12.85} & \textbf{13.90} & \textbf{8.44} & \textbf{4.80} & \textbf{8.21} \\
\hline
\end{tabular}}
\end{table}

\paragraph{Adaptation with reduced paired speech.}
To isolate the contribution of text-first initialization, we compare twosystems under an identical paired-speech budget: a single real-posterior LoRA epoch on about one ninth of the complete MLS training set. TASU-Real-LoRA (one epoch) starts from a randomly initialized projector and the original Qwen backbone, whereas TASU-CPS$\rightarrow$Real-LoRA starts from the CPS-trained checkpoint. As shown in Table~\ref{tab:lora_results}, CPS initialization reduces pooled WER from 11.04 to 9.19, a 16.8\% relative reduction. The largest absolute gains occur on the three lower-resource languages, Polish (26.76$\rightarrow$18.32), Italian (17.78$\rightarrow$14.85), and Portuguese (12.25$\rightarrow$9.47), for which the reduced subset provides the fewest paired utterances. These results show that multilingual alignment can first be established from abundant transcripts and then adapted into an effective recognizer with only a small amount of paired-speech supervision.

\paragraph{Simulator language-coverage ablation.}
Table~\ref{tab:coverage_ablation} examines whether the learned posterior simulator can generalize to a language excluded from its own training data. For this ablation, we train the simulator on seven languages by removing Polish teacher-posterior pairs. We then use this seven-language simulator to generate CTC posteriors for transcripts from all eight languages, including Polish. The downstream TASU model is therefore still trained on simulated posteriors for all eight languages; Polish is excluded only from simulator training, not from TASU training.

Removing Polish from simulator training has little effect on the seven languages seen by the simulator: their macro-average WER changes from 10.79 to 10.61. However, Polish WER increases from 16.38 to 80.47. This degradation cannot be attributed to missing Polish transcripts during TASU training. Instead, it shows that the simulator does not generate useful Polish CTC posteriors when Polish is absent from its teacher-posterior training data, and the downstream projector and LoRA parameters cannot compensate for this mismatch. The ablation therefore suggests that target-language coverage is important at the simulator-training stage, even when TASU itself is trained on text from all target languages.

\begin{table}[t]
\centering
\caption{Effect of removing Polish from simulator training. Both TASU models are trained on all eight languages; Non-PL macro is the macro-average WER over the other seven.}
\label{tab:coverage_ablation}
\begin{tabular}{lrrr}
\hline
Simulator training & Pl & Non-PL macro & Overall \\
\hline
8-language simulator & 16.38 & 10.79 & 10.51 \\
7-language simulator (w/o Polish) & 80.47 & 10.61 & 12.02 \\
\hline
\end{tabular}
\end{table}

Taken together, the experiments distinguish two requirements for extending posterior-based alignment beyond English. First, a single shared TASU model can refine real multilingual CTC posteriors consistently, as demonstrated by
the across-language gains of TASU-Real-LoRA. Second, text-derived posterior training is constrained by the mismatch between simulated training posteriors and real inference posteriors, as well as by language coverage during simulator training. The single Polish removal experiment should not be interpreted as proving full independence among all language pairs. Nevertheless, the result shows that a simulator does not necessarily transfer to a language absent from its teacher-posterior data, even when downstream TASU is trained on that language's transcripts.

To determine whether these observations are specific to the OWSM-CTC encoder, we next replicate the core training configurations with an alternative Whisper-CTC frontend.

\subsection{Generalization to Alternative CTC Encoders}
\label{subsec:whisper_ctc}

The preceding experiments establish TASU with OWSM-CTC as the posterior
provider. To verify that the observed gains are not specific to this
encoder, we replicate the key training configurations with an alternative
Whisper-CTC frontend. This encoder is derived from Whisper-Large-v3 with
a randomly initialized CTC head trained on a limited multilingual subset,
resulting in a different posterior distribution and substantially higher
baseline WER. Despite this mismatch in absolute performance, the relative
behavior of the three TASU configurations remains consistent.

\begin{table}[t]
\centering
\caption{WER (\%) with a Whisper-CTC frontend. All TASU conditions use the 150k-vocab CTC head. Phase1$\to$Phase2 uses text initialization followed by audio-derived real-posterior adaptation. Nl, En, Fr, De, It, Pl, Pt, and Es denote Dutch, English, French, German, Italian, Polish, Portuguese, and Spanish, respectively; Overall denotes pooled WER.}
\label{tab:whisper_ctc}
\resizebox{\textwidth}{!}{%
\begin{tabular}{@{}lrrrrrrrrr@{}}
\hline
System & Nl & En & Fr & De & It & Pl & Pt & Es & Overall \\
\hline
Whisper-CTC (150k vocab) & 29.51 & 17.18 & 26.67 & 26.01 & 37.57 & 51.67 & 36.33 & 20.24 & 25.91 \\
Whisper-CTC (38k vocab)  & 28.45 & 16.73 & 24.57 & 24.04 & 35.09 & 47.42 & 34.30 & 18.85 & 24.42 \\
\hline
TASU (audio only)        & 20.94 & 14.35 & 14.60 & 17.44 & 25.88 & 36.81 & 22.29 & 12.50 & 17.62 \\
TASU (CPS, Phase1 only) & 42.15 & 22.36 & 32.43 & 36.87 & 52.80 & 74.83 & 47.97 & 29.35 & 35.54 \\
TASU (CPS, Phase1$\to$2)& \textbf{18.70} & \textbf{12.82} & \textbf{11.84} & \textbf{13.79} & \textbf{21.09} & \textbf{29.45} & \textbf{17.89} & \textbf{9.19} & \textbf{14.67} \\
\hline
\end{tabular}}
\end{table}

Table~\ref{tab:whisper_ctc} shows that the two-stage
Phase1$\rightarrow$Phase2 curriculum transfers effectively to the
Whisper-CTC frontend. Compared with standalone Whisper-CTC, it reduces
WER in all eight languages. It also outperforms audio-only TASU in seven
languages, with Dutch as the only exception, and reduces pooled WER from
17.62 to 14.67, corresponding to a 16.7\% relative reduction.

Phase1 alone performs poorly because the model is trained with
text-derived posteriors but evaluated with real Whisper-CTC posteriors.
After one epoch of real-posterior adaptation, however, its pooled WER
decreases from 35.54 to 14.67. Together with the OWSM-CTC results, this
shows that text-first initialization can be effectively adapted to real
posterior inputs with different CTC frontends, although the exact ranking
between audio-only and two-stage training depends on the posterior
provider and adaptation setting.

The reduced-vocabulary Whisper-CTC head improves the standalone pooled
WER from 25.91 to 24.42. Because downstream TASU experiments use only the
full-vocabulary head, this comparison is reported as a frontend diagnostic
rather than evidence about the effect of vocabulary size on TASU.

In summary, the Whisper-CTC replication validates that multilingual
TASU generalizes across different CTC encoders. The consistent relative
improvements from curriculum learning, together with the preserved
ranking of training configurations, support TASU as a general
multilingual alignment framework rather than a method tailored to a
specific acoustic posterior provider.

\section{Conclusion and Future Work}
\label{sec:conclusion}

This study demonstrates that TASU can be extended from English to a single shared multilingual Speech LLM through the native OWSM-CTC posterior space. Adapting the projector and language model with real posteriors improves recognition consistently across all eight languages, while posteriors generated from text provide a practical initialization for subsequent speech adaptation. In particular, the learned multilingual simulator substantially improves over heuristic CPS, and a text-initialized TASU model can be efficiently calibrated with a limited amount of real-posterior supervision. These results support a practical multilingual training paradigm in which abundant text first establishes the posterior-to-LLM alignment and a smaller amount of paired speech subsequently adapts it to real acoustic posteriors.

Future work will extend this framework in two directions. First, we will unfreeze the CTC head during TASU training, allowing the acoustic posterior space to adapt jointly with the projector and language model rather than remaining fixed. This is motivated by the
Whisper-CTC experiments in Section~\ref{subsec:whisper_ctc}, where a
weak CTC head limited the absolute performance ceiling despite strong
relative gains from TASU alignment. Second, we will use the multilingual posterior simulator to generate additional training supervision from text for low-resource languages, reducing language imbalance without requiring corresponding paired speech. 

{\small
\subsubsection{\ackname} This work has been supported by the China NSFC Project (No. 92370206), the Key Research and Development Program of Jiangsu Province, China (Grant No. BF2025029), and the Joint Fund Project of Shanghai Commercial Aircraft Systems Engineering Science and Innovation Center Project (No. B2025JS36236).

\subsubsection{Disclosure of Interests} The authors have no competing interests to declare that are relevant to the content of this article.
}
\newpage

\begin{thebibliography}{99}

\bibitem{arora2025landscape}
Siddhant Arora, Kai-Wei Chang, Chung-Ming Chien, Yifan Peng, Haibin Wu, Yossi Adi, Emmanuel Dupoux, Hung-Yi Lee, Karen Livescu, and Shinji Watanabe, ``On the landscape of spoken language models: A comprehensive survey,'' \emph{arXiv preprint arXiv:2504.08528}, 2025.

\bibitem{peng2024survey}
Jing Peng, Yucheng Wang, Bohan Li, Yiwei Guo, Hankun Wang, Yangui Fang, Yu Xi, Haoyu Li, Xu Li, Ke Zhang, Shuai Wang, and Kai Yu, ``A survey on speech large language models for understanding,'' \emph{IEEE Journal of Selected Topics in Signal Processing}, vol. 20, no. 1, pp. 2--31, 2026. doi: 10.1109/JSTSP.2025.3640535.

\bibitem{radford2023whisper}
Alec Radford, Jong Wook Kim, Tao Xu, Greg Brockman, Christine McLeavey, and Ilya Sutskever, ``Robust speech recognition via large-scale weak supervision,'' in \emph{Proc. ICML}, 2023, pp. 28492--28518.

\bibitem{tang2023salmonn}
Changli Tang, Wenyi Yu, Guangzhi Sun, Xianzhao Chen, Tian Tan, Wei Li, Lu Lu, Zejun Ma, and Chao Zhang, ``SALMONN: Towards generic hearing abilities for large language models,'' in \emph{Proc. ICLR}, 2024.

\bibitem{chu2024qwen2}
Yunfei Chu, Jin Xu, Qian Yang, Haojie Wei, Xipin Wei, Zhifang Guo, Yichong Leng, Yuanjun Lv, Jinzheng He, Junyang Lin, et al., ``Qwen2-Audio technical report,'' \emph{arXiv preprint arXiv:2407.10759}, 2024.

\bibitem{ma2025legoslm}
Rao Ma, Tongzhou Chen, Kartik Audhkhasi, and Bhuvana Ramabhadran, ``LegoSLM: Connecting LLM with speech encoder using CTC posteriors,'' \emph{arXiv preprint arXiv:2505.11352}, 2025.

\bibitem{ma2024embarrassingly}
Ziyang Ma, Guanrou Yang, Yifan Yang, Zhifu Gao, Jiaming Wang, Zhihao Du, Fan Yu, Qian Chen, Siqi Zheng, Shiliang Zhang, et al., ``An embarrassingly simple approach for LLM with strong ASR capacity,'' \emph{arXiv preprint arXiv:2402.08846}, 2024.

\bibitem{xu2025qwen2}
Jin Xu, Zhifang Guo, Jinzheng He, Hangrui Hu, Ting He, Shuai Bai, Keqin Chen, Jialin Wang, Yang Fan, Kai Dang, et al., ``Qwen2.5-Omni technical report,'' \emph{arXiv preprint arXiv:2503.20215}, 2025.

\bibitem{ding2025kimi}
KimiTeam, Ding Ding, Zeqian Ju, Yichong Leng, Songxiang Liu, Tong Liu, Zeyu Shang, Kai Shen, Wei Song, Xu Tan, Heyi Tang, et al., ``Kimi-Audio technical report,'' \emph{arXiv preprint arXiv:2504.18425}, 2025.

\bibitem{peng2026tasu}
Jing Peng, Yi Yang, Xu Li, Yu Xi, Quanwei Tang, Yangui Fang, Junjie Li, and Kai Yu, ``TASU: Text-only alignment for speech understanding,'' \emph{arXiv preprint arXiv:2511.03310}, 2025.

\bibitem{peng2026tasu2}
Jing Peng, Chenghao Wang, Yi Yang, Lirong Qian, Junjie Li, Yu Xi, Shuai Wang, and Kai Yu, ``TASU2: Controllable CTC simulation for alignment and low-resource adaptation of speech LLMs,'' \emph{arXiv preprint arXiv:2604.08384}, 2026.

\bibitem{graves2006connectionist}
Alex Graves, Santiago Fernández, Faustino Gomez, and J{\"u}rgen Schmidhuber, ``Connectionist temporal classification: Labelling unsegmented sequence data with recurrent neural networks,'' in \emph{Proceedings of the 23rd International Conference on Machine Learning}, 2006, pp. 369--376.

\bibitem{chen2016phone}
Zhehuai Chen, Wei Deng, Tao Xu, and Kai Yu, ``Phone synchronous decoding with CTC lattice,'' in \emph{Proc. Interspeech}, 2016, pp. 1923--1927.

\bibitem{chen2017ctc}
Zhehuai Chen, Yimeng Zhuang, Yanmin Qian, and Kai Yu, ``Phone synchronous speech recognition with CTC lattices,'' \emph{IEEE/ACM Transactions on Audio, Speech, and Language Processing}, vol. 25, no. 1, pp. 90--101, 2017. doi: 10.1109/TASLP.2016.2625459.

\bibitem{chen2019lsd}
Zhehuai Chen, Wenlu Zheng, Yongbin You, Yanmin Qian, and Kai Yu, ``Label Synchronous Decoding for Speech Recognition,'' \emph{Chinese Journal of Computers}, vol. 42, no. 7, pp. 1511--1523, 2019.

\bibitem{peng2024owsmctc}
Yifan Peng, Yui Sudo, Muhammad Shakeel, and Shinji Watanabe, ``OWSM-CTC: An open encoder-only speech foundation model for speech recognition, translation, and language identification,'' in \emph{Proc. ACL}, 2024, pp. 10192--10209.

\bibitem{pratap2020mls}
Vineel Pratap, Qiantong Xu, Anuroop Sriram, Gabriel Synnaeve, and Ronan Collobert, ``MLS: A large-scale multilingual dataset for speech research,'' in \emph{Proc. Interspeech}, 2020, pp. 2757--2761. doi: 10.21437/Interspeech.2020-2826.

\bibitem{zhang2022wenet}
Binbin Zhang, Di Wu, Zhendong Peng, Xingchen Song, Zhuoyuan Yao, Hang Lv, Lei Xie, Chao Yang, Fuping Pan, and Jianwei Niu, ``WeNet 2.0: More productive end-to-end speech recognition toolkit,'' \emph{arXiv preprint arXiv:2203.15455}, 2022.

\bibitem{qwen25}
An Yang et al., ``Qwen2.5 technical report,'' \emph{arXiv preprint arXiv:2412.15115}, 2024.

\bibitem{hu2022lora}
Edward J. Hu, Yelong Shen, Phillip Wallis, Zeyuan Allen-Zhu, Yuanzhi Li, Shean Wang, Lu Wang, and Weizhu Chen, ``LoRA: Low-rank adaptation of large language models,'' in \emph{Proc. ICLR}, 2022.

\bibitem{loshchilov2019adamw}
Ilya Loshchilov and Frank Hutter, ``Decoupled weight decay regularization,'' in \emph{Proc. ICLR}, 2019.

\bibitem{rajbhandari2020zero}
Samyam Rajbhandari, Jeff Rasley, Olatunji Ruwase, and Yuxiong He, ``ZeRO: Memory optimizations toward training trillion parameter models,'' in \emph{Proc. SC}, 2020, Art. no. 20. doi: 10.1109/SC41405.2020.00024.

\end{thebibliography}
%
\bibliographystyle{splncs04}

\end{document}